\documentclass{moriondc}

\def\be{\begin{equation}}
\def\ee{\end{equation}}
\def\bea{\begin{eqnarray}}
\def\eea{\end{eqnarray}}

\begin{document}
\vspace*{4cm}
\title{Backreaction effects on the matter side of Einstein's field equations}

\author{S. Floerchinger$^{1}$, N. Tetradis$^{2}$ and U.A.Wiedemann$^{1}$}
\address{$^{1}$ Physics Department, Theory Unit, CERN, CH-1211 Gen\`eve 23, Switzerland\\
$^{2}$ Department of Physics, University of Athens, Zographou 157 84, Greece}

% \author{S. Floerchinger$^{1}$, N. Tetradis$^{2}$ and \underline{U.A.Wiedemann}$^{1}$}

% \address{$^{1}$ Physics Department, Theory Unit, CERN, CH-1211 Gen\`eve 23, Switzerland\\
% $^{2}$ Department of Physics, University of Athens, Zographou 157 84, Greece}

\maketitle\abstracts{
Recently, we have derived a novel and compact expression for how perturbations in the matter fields 
of the cosmological fluid can lead to deviations from the standard Friedmann equations. Remarkably, 
the dissipative damping of velocity perturbations by bulk and shear viscosity in the dark sector can
modify the expansion history of the universe on arbitrarily large scales. In universes in which this effect
is sufficiently sizeable, it could account for the acceleration of the cosmological 
expansion. But even if dark matter should be less viscous and if the effect would be correspondingly smaller, 
it may have observable consequences in the era of precision cosmology. Here, we review the origin of this backreaction
effect and possibilities to constrain it further. }

\section{Backreaction effects in cosmology}

In the cosmological perturbation theory that is at the basis of the phenomenologically successful $\Lambda$CDM
model, all fields entering Einstein's field equations are split into a spatially homogeneous and isotropic {\it background} part 
and {\it perturbations} around it. For instance, one writes 
$g_{\mu\nu}(x) = \bar g_{\mu\nu}(\tau) + \delta g_{\mu\nu}(x)$ for the metric, $\epsilon(x) = \bar \epsilon(\tau) +
\delta \epsilon(x)$ for the energy density, etc., where the background fields are defined
as spatial averages of the full fields. Up to first order in perturbations, the spatially averaged Einstein equations reduce then to the well-known Friedmann
equations for the background fields. However, Einstein's field equations are non-linear, and, beyond first order in perturbations, a spatial 
average of Einstein's equations for the full fields differs from Einstein's field equations for the spatially averaged fields. 
The perturbations {\it backreact}, that is,  
the time evolution of the background fields depends on the time evolution of the perturbations. 

The possible sources of such backreaction effects, and the question to what extent they may modify Friedmann's equations have been 
discussed repeatedly. On the one hand, it has been argued that the formation of non-linear structures could affect the
average expansion rate of the Universe considerably and that this may even account for the accelerated expansion in late-time cosmology, for a 
review see Ref.~\cite{Buchert:2011sx}. In sharp contrast to this line of thought, there are also arguments that the phenomenological success of the 
$\Lambda$CDM model is a consequence of its mathematical consistency, in the sense that the evolution of the exactly
homogeneous and isotropic FLRW metric does indeed account with high precision for the evolution of spatially averaged fields in our
physically realized Universe with perturbations, for a review see Ref.~\cite{Green:2014aga}. So far, this discussion focussed mainly
on backreactions of metric perturbations, that is, terms in Einstein's field equations that are second order or higher in $\delta g_{\mu\nu}$. 
Instead, we point here to backreaction effects that are second order in matter perturbations and that have
a physical interpretation in terms of dissipative processes.

\section{Backreaction from matter perturbations and cosmological expansion}
\label{sec2}
 Einstein's equation for an exactly homogeneous and isotropic FLRW universe are the first Friedmann equation, 
% $\epsilon = (1/8\pi G_N) H^2$ 
that relates $\bar\epsilon$ to the Hubble parameter $H$, and the second Friedmann equation.
The only additional information contained in the latter is energy conservation for an expanding system,
\begin{equation}
\frac{1}{a}\dot{\bar\epsilon} + 3 H \left(\bar\epsilon + \bar p - 3\zeta H\right) = D \, ,
\label{eq1}
\end{equation}
where $D=0$ for an unperturbed FLRW metric, $H = \dot a/a^2$, and the dot denotes a derivative w.r.t. conformal time.
We recall these elementary facts to highlight two issues that are important for the following. First, what enters the
evolution of the background fields is not the average pressure $\bar p$, but the effective pressure $\bar p_{\rm eff} = \bar p - 3\zeta H$ 
for which a term proportional to the bulk viscosity $\zeta$ times expansion scalar is subtracted. Indeed, 
bulk viscosity is the only dissipative effect that can arise in an exactly homogeneous and isotropic system, and bulk viscous pressure is
a negative pressure. The logical possibility that bulk viscosity may therefore account for phenomena normally attributed to dark energy 
has been explored in several studies~\cite{Li:2009mf,Gagnon:2011id}. Second, equation
(\ref{eq1}) illustrates the well-known fact that Einstein's field equations include the conditions for energy momentum conservation, 
$\nabla_\mu T^{\mu\nu} = 0$. In our recent study~\cite{Floerchinger:2014jsa}, we started from energy conservation. 
Working in the Landau Frame  $-u_\mu T^{\mu\nu} \equiv \epsilon u^\nu$ and keeping shear viscous and bulk viscous effects up to 
first order in a gradient expansion of the fluid dynamic fields, one finds from $\nabla_\mu T^{\mu 0} = 0$
\begin{equation}
u^\mu \partial_\mu \epsilon + (\epsilon + p ) \nabla_\mu u^\mu   - \zeta \Theta^2 - 2 \eta \sigma^{\mu\nu} \sigma_{\mu\nu} = 0\, ,
\label{eq2}
\end{equation}
where $\eta$ and $\zeta$ denote the shear and bulk viscosity, respectively, $\Delta^{\mu\nu} = g^{\mu\nu} + u^\mu u^\nu$ and  
\begin{eqnarray}
 \sigma^{\mu\nu} &=&  \frac{1}{2} \left[ \Delta^{\mu\alpha} \Delta^{\nu\beta} +  \Delta^{\mu\beta} \Delta^{\nu\alpha} - \frac{2}{3} \Delta^{\mu\nu} \Delta^{\alpha\beta} \right] \nabla_\alpha u_\beta\, , 
\nonumber\\
 \Theta  &=&  \nabla_\mu u^\mu\, .
\label{eq3}
\end{eqnarray}
Forming the spatial average of eq.~(\ref{eq2}), one can determine the dissipative corrections to the time evolution of the average energy density $\bar\epsilon$.
In general, metric perturbations enter this spatial average via the covariant derivative $\nabla_\mu$, the projector $\Delta_{\mu\nu}$ and the normalization
of the velocity field. For instance, in conformal Newtonian gauge with Newtonian potentials $\Psi$ and $\Phi$, the constraint $u_\mu u^\mu = -1$ implies 
$u^\mu =\left(\gamma, \gamma \vec v\right)$, where $\gamma = 1/(a\sqrt{1-\vec v^2+2\Psi+2\Phi \vec v^2})$. However, in the relevant limit of
small gravitational potentials ($\Phi\, ,\Psi \ll 1$), that vary slowly in time, $\dot\Phi \sim \textstyle\frac{\dot a}{a} \Phi$, one finds that the spatial
average of (\ref{eq2}) is independent of metric perturbations~\cite{Floerchinger:2014jsa}. On the other hand, there are fluid perturbations. In particular,
perturbations in the gradients of the velocity field enter the terms $\Theta^2$ and $\sigma^{\mu\nu} \sigma_{\mu\nu}$
in eq.~(\ref{eq2}) to second order, and thus they do not vanish upon spatial averaging. Since $\Theta^2$ and $\sigma^{\mu\nu} \sigma_{\mu\nu}$
are positive semidefinite, one finds (for small fluid velocity $v$) that  the spatial average of (\ref{eq2}) is of the form of (\ref{eq1}) with a positive 
term~\cite{Floerchinger:2014jsa} 
\begin{equation}
D=  \textstyle\frac{1}{a^2} \langle \eta \left[ \partial_i v_j \partial_i v_j + \partial_i v_j \partial_j v_i - \textstyle\frac{2}{3} \partial_i v_i \partial_j v_j\right] \rangle
 + \textstyle\frac{1}{a^2} \langle \zeta[\vec \nabla \cdot \vec v]^2 \rangle +
\textstyle\frac{1}{a}\langle  \vec v \cdot \vec \nabla \left( p - 6 \zeta H \right)\rangle \, .
\label{eq4}
\end{equation}
Here, the term $D$ gives an explicit backreaction of fluid perturbations on the time evolution of the average energy density $\bar \epsilon$. In Ref.~\cite{Floerchinger:2014jsa},
we have analyzed the effect of this term on the cosmological expansion. To this end, we combined eq.(\ref{eq1}) with an equation for the scale factor $a$. For the latter, we used
the trace of Einstein's equation since its spatial average does not depend on fluid velocity perturbations,
\begin{equation}
	R = 8\pi G_{\rm N} T_\mu^\mu = - 8\pi G_{\rm N}  \left( \bar\epsilon - 3 \bar p_{\rm eff} \right) % = 6 \left(\frac{1}{a}\dot H + 2 H^2 \right) 
            = - 6\frac{\ddot a}{a^3} \, .
         \label{eq5}
\end{equation}
 Combining both equations, one finds for the deceleration parameter 
$q=-1-\dot H/ (a H^2)$ the differential equation~\cite{Floerchinger:2014jsa}  (we use the simple equation of state $\bar p_{\rm eff} = \hat w\, \bar\epsilon$)
\begin{equation}
-\frac{d q}{d \ln a} + 2 (q-1) \left(q-\textstyle\frac{1}{2}(1+3 \hat w)\right) = \frac{4\pi G_{\rm N} D(1-3\hat w)}{3 H^3}.
\label{eq6}
\end{equation}
The fixed point of this equation is accelerating, i.e. $q < 0$, for $\frac{4\pi G_{\rm N} D}{3 H^3} > \frac{1+3\hat w}{1-3\hat w}$.

\section{An explicit deviation from Friedmann's equation for the (toy) ansatz $D \propto H$}
\label{sec3}
The conclusion of the discussion above is that in a universe in which dissipative effects are sufficiently large, $D \sim O(H^3/G_{\rm N})$, dissipation induces accelerated 
cosmological expansion. Clearly, in a universe in which dissipative effects are smaller, other physical effects need to be invoked to account for an observed accelerated
expansion. However, even for small but non-vanishing backreaction $D$, the cosmic expansion scale factor $a$ will not obey the exact Friedmann equation. We emphasize
this point since it is a key assumption of the $\Lambda$CDM cosmological concordance model\cite{Bertschinger:2006aw} that $a$ does satisfy the Friedmann equation; 
this assumption underlies for instance the definition of cosmological distance measures and it is poorly tested at best. 
Given the precision of modern cosmology, it is thus interesting to explore the sensitivity of distance measures on $D$ by analyzing deviations from Friedmann's
equation induced by $D$. 

The term $D$ is calculable once the matter content of the Universe and the dissipative properties of dark matter are specified. $D$ affects the time evolution of the background fields
in eq.~(\ref{eq1}), while the cosmological perturbations needed to determine the spatial average $D$ according to eq.~(\ref{eq4}) are evolved on top of these background fields. 
This complicates a dynamical determination of  $D$. 
In general, since perturbations evolve, $D$ will have a non-trivial dependence on redshift, $D=D(z)$, where $a = \textstyle\frac{1}{1+z}$.  We plan to calculate this evolution of $D$ in 
future work for different viscous properties of the dark sector~\cite{workinprogress}. For the purpose of these proceedings, we just examine heuristically one type of $z$-dependence of $D$:

Let us consider 
for a moment a hypothetical universe composed of radiation and viscous dark matter, but without dark energy. What would we have to choose for the dissipative backreaction term $D$ of that
hypothetical universe, if we required that its deceleration parameter matches that of the phenomenologically successful $\Lambda$CDM universe? To answer this question, we 
write the scale dependence of the Hubble parameter of the $\Lambda$CDM universe, $H(z) = H_0 \sqrt{\Omega_\Lambda + \Omega_M (1+z)^3 + \Omega_R (1+z)^4}$, 
we derive from it the explicit form for the deceleration parameter $q(z)$ in the $\Lambda$CDM universe, we insert it into the left hand side of equation (\ref{eq6}) and we read off
an expression for $D$. For the case of a pressure-less fluid $\hat w = 0$, we find $D = \textstyle\frac{3}{4\pi G_{\rm N}} 6 \Omega_\Lambda H_0^2 H(z)$. Therefore, if one would
aim at exploring the possibility that dissipative properties of dark matter can account for phenomena normally attributed to dark energy, one would seek dark matter with viscous 
properties that lead to a backreaction $D(z) \approx {\rm const}\cdot H(z)$. 

We caution that by making the {\it ad hoc} ansatz $D(z) = {\rm const}\cdot H(z)$, we did not check whether this particular scale dependence can arise from a dynamical
evolution of perturbations in a fluid with 'realistic' viscous properties. In general, both the size and the scale dependence of $D(z)$ depend on the viscous properties of dark 
matter. There is no {\it a priori} reason that the scale dependence satisfies the simple ansatz $D(z) = {\rm const}\cdot H(z)$. For a more general form of $D(z)$, the 
solution of the evolution equations for the background field $\bar\epsilon$ will show deviations from Friedmann's equation that involve an integration over the
evolution of $D$. However, for the ansatz $D(z) = {\rm const}\cdot H(z)$, a linear superposition of eqs.~(\ref{eq1}) and (\ref{eq5}) results in
%  the deviations 
% from the first and second standard Friedmann equation take a particularly simple, algebraic form, 
%
\begin{equation}
\bar \epsilon - \frac{D}{4H} = % \hat\epsilon +  \frac{D}{12 H} 
\frac{3}{8\pi G_{\rm N}} H^2\, ,\quad \hbox{and} \quad  
\bar p_{\rm eff} - \frac{D}{12 H} =  -  \frac{1}{8\pi G_{\rm N}} \left(2 \textstyle\frac{1}{a} \dot H + 3 H^2 \right)\, .
\label{eq7}
\end{equation}
Here, the relation between the Hubble parameter and the total energy density $\bar\epsilon$ and total effective pressure differ from the standard Friedmann equations
by the terms proportional to $D$. There is a one-to-one relation between this type of evolution and a $\Lambda$CDM model with energy density 
$\hat{\epsilon} = \bar\epsilon - \textstyle\frac{D}{3H}$ and dark energy density $\frac{D}{12 H} $. More precisely, in terms of $\hat{\epsilon} $, the evolution equation 
(\ref{eq1}) for energy density, 
and the trace (\ref{eq5}) of Einstein's equations can be written  as
\begin{equation}
	\frac{1}{a}\dot{\hat{\epsilon}} + 3 H \left(\hat{\epsilon} + p_{\rm eff} \right) = 0\, ,
	\quad \hbox{and} \quad
	 R + \frac{8\pi G_{\rm N}\, D}{3\, H} = - 8\pi G_{\rm N}  \left( \hat{\epsilon} - 3 \bar p_{\rm eff} \right)\, .
	\label{eq8}
\end{equation}
This is of the same form as the corresponding equations in the $\Lambda$CDM model if one specializes to a pressureless fluid $\bar p_{\rm eff}=0$ and if one
identifies $\hat{\epsilon}$ with the energy density in the $\Lambda$CDM model and the cosmological constant with 
\begin{equation}
 \Lambda = \frac{2\pi G_{\rm N} D}{3 H}\, .
\label{eq9}
\end{equation}

\section{Outlook}
Rather than repeating or anticipating work published in refereed journals, we have made use of this proceedings article to share some simple but heuristic 
considerations of how dissipative backreaction of matter perturbations may lead to deviations from Friedmann's equation
for the scale parameter. Clearly,  a more rigorous analysis of these deviations is possible: one can calculate the size and scale dependence of $D(z)$
as a function of the viscous properties of dark matter. This calculation could contribute to a critical appraisal of cosmological distance measures. 
This line of thought will be presented elsewhere. 

We finally remark on a crucial point that was discussed more prominently in the oral presentation of our work in Moriond.
The evolution of perturbations and the growth of cosmological structure depends on dissipative properties, and information about structure formation is known 
to constrain possible dissipative phenomena~\cite{Li:2009mf,Gagnon:2011id}. To first order in cosmological perturbations, one finds
from Einstein equations that the evolution equation of the density contrast $\delta \rho = \delta \epsilon/\bar{\epsilon}$ does not depend explicitly on viscous contributions,
while the evolution of fluid velocity perturbations is attenuated by a scale-dependent viscous term. As a consequence, one expects that unlike finite pressure
that counteracts the growth of density contrast, finite viscosity will not counteract this growth but it will slow down the growth of velocity gradients and it may thus
delay structure formation. We are currently evaluating~\cite{workinprogress} how the constraints on viscous properties arising from structure formation translate into 
constraints on the backreaction $D$.

\end{document}